# Stochastic dynamical theory of power-law distributions induced by multiplicative noise

Du Jiulin

*Department of Physics, School of Science, Tianjin University, Tianjin 300072, China*

The two-variable Langevin equations, modeling the Brownian motion of a particle moving in a potential and leading to the Maxwell-Boltzmann distribution of the corresponding Fokker-Planck equation, are shown to give rise to types of stationary power-law distributions through the multiplicative noise. The power-law distributions induced by this inhomogeneous noise are proved to be a result of that the relation of diffusion to friction depends on the energy. We understand the conditions under which the power-law distributions are produced and how they are produced in systems away from equilibrium, and hence derive a generalized fluctuation-dissipation theorem. This leads to a generalized Klein-Kramers equation, and a generalized Smoluchowski equation for the particle moving in a strong friction medium, whose stationary-state solutions are exactly Tsallis distribution.

There are a variety of numerous non-exponential or power-law distributions that have been frequently observed and studied in the systems away from equilibrium. The typical forms of such power-law distributions that have been widely paid attention to contain the $\kappa$-distributions or the generalized Lorentzian distributions in the solar wind and space plasmas [1-6], the $q$-distributions in complex systems within Tsallis statistics [7-13], and those noted in physics, chemistry, and elsewhere like $P(E) \sim E^{-\alpha}$ with an index $\alpha >0$ [2, 10,14-17]. Simultaneously, a type of statistical mechanical theory of studying the power-law distributions is being constructed via generalizing Gibbsian theory [18] to systems away from thermal equilibrium or through generalizing Boltzmann entropy to Tsallis entropy [7] for complex systems, etc. But, the details of stochastically dynamical origins about the underlying power-law distributions are mostly unknown. An open problem, which is being investigated intensively, is the conditions under which they are produced and how they are produced in systems away from equilibrium [8, 9,19], which goes beyond the realm governed by the traditional statistical theory with Boltzmann-Gibbs distribution.

The stochastic dynamics to give rise to the Maxwell-Boltzmann distribution could be clearly explained by the two-variable Langevin equations [20, 21], modeling the Brownian motion of a particle moving in a potential field and under the influence of a friction force and a white noise. The correlation strength of the noise, usually referred to as diffusion coefficient, is a constant, associated with a friction coefficient by the fluctuation-dissipation theorem. In this letter, it is shown that this simple and elegant landscape would reappear when we are accounting for the microscopic dynamical origins to give rise to types of power-law distributions. We introduce the multiplicative noise with inhomogeneous correlation strength into the Langevin equations. An energy-dependent relation of diffusion to friction is determined by the way to solve the corresponding stationary Fokker-Planck (F-P) equation. We hence easily derive a generalized fluctuation-dissipation theorem, one condition under which we could understand the microscopic dynamical origins giving rise to a type of power-law distributions.

Let us consider the two-variable Langevin equations for ($x$, $p$), modeling the Brownian motion of a particle moving in a potential field $V(x)$,

$$\frac{dx}{dt} = \frac{p}{m}, \tag{1}$$

$$\frac{dp}{dt} = -\frac{dV(x)}{dx} - \gamma\, p + \eta(x,p,t), \tag{2}$$

where the friction force is $-\gamma p$. Generally speaking, the friction coefficient may be considered as a function of $x$ and $p$, i.e. $\gamma = \gamma(x,p)$, if the Brownian particle moves in an inhomogeneous complex medium. The noise $\eta(x,p,t)$ is now defined as multiplicative and Gaussian, with zero average and the delta-correlated in time $t$,

$$\langle \eta(x,p,t) \rangle = 0, \quad \langle \eta(x,p,t)\eta(x,p,t') \rangle = 2D(x,p)\delta(t-t'). \tag{3}$$

So the correlation strength of the noise (i.e. diffusion coefficient), $D(x,p)$, instead of a constant, is regarded as a function of $x$ and $p$. Such a multiplicative noise is an inhomogeneous one.

If $\rho(x, p, t)$ is the probability distribution function of variables $(x, p)$ at time $t$, the corresponding F-P equation to the above Langevin equations [20] is

$$\frac{\partial \rho}{\partial t} = -\frac{p}{m}\frac{\partial \rho}{\partial x} + \frac{\partial}{\partial p}\left(\frac{dV}{dx} + \gamma\, p\right)\rho + \frac{\partial}{\partial p}\left(D\frac{\partial \rho}{\partial p}\right). \tag{4}$$

The stationary F-P equation satisfies

$$-\frac{p}{m}\frac{\partial \rho_s}{\partial x} + \frac{\partial}{\partial p}\left(\frac{dV}{dx} + \gamma\, p\right)\rho_s + \frac{\partial}{\partial p}\left(D\frac{\partial \rho_s}{\partial p}\right) = 0. \tag{5}$$

As a macroscopic distribution of a system, we seek a stationary-state solution of this equation in the form of $\rho_s(x, p) = \rho_s(E(x, p))$, with the energy: $E(x, p) = V(x) + p^2/2m$. Then, from Eq.(5) we have

$$\frac{\partial}{\partial p}\left(\gamma\, p\rho_s + \frac{p}{m}D\frac{d\rho_s}{dE}\right) = 0. \tag{6}$$

By solving this differential equation, we derive that

$$\rho_s(E) = Z^{-1}\exp\left(-m\int\frac{\gamma}{D}dE\right)\cdot\exp\left(\int\frac{mC(x)}{pD\rho_s(E)}dE\right), \tag{7}$$

where $Z$ is a normalization constant; $C(x)$ is the "integral constant" as a result of performing the integral on $p$, so it can be any function of $x$. However, as we know, the Langevin equations give rise to the Maxwell-Boltzmann distribution as the stationary solution of the corresponding Fokker-Planck equation if the system reaches a thermal equilibrium at a long time, and the correlation strength of the noise as well as the friction coefficient are also constants, a condition under which the fluctuation-dissipation theorem states that $D = m\gamma\beta^{-1}$, with $\beta = 1/kT$. Hereby the "integral constant" in Eq.(7) is discarded, $C(x) = 0$. And then, the stationary-state solution reads

$$\rho_s(E) = Z^{-1}\exp\left(-m\int\frac{\gamma}{D}dE\right). \tag{8}$$

It is clear that Eq.(8) is the Maxwell-Boltzmann distribution if there is the fluctuation-dissipation relation: $D/m\gamma = \beta^{-1}$. Now let us consider $(D/m\gamma)$ as a function of the energy, i.e. $D/m\gamma = f(E)$, and if $f(E)$ is a continuous and differentiable function at least in the surrounding of $E=0$, then it can be expressed as the Maclaurin series, i.e., $f(E) = f(0) + f'(0)E + \cdots + f^{(n)}(0)E^n/n! + \cdots$. Let $f(0) = \beta^{-1}$, we find that if $f'(E) \approx$ const., or if the energy $E$ is small, then the terms of $n \geq 2$ in the series can be neglected; hence we have $f(E) \approx \beta^{-1} + f'(0)E$, and

$$D = m\gamma\beta^{-1}(1 - \kappa\beta E), \tag{9}$$

with a parameter $\kappa \equiv -f'(0)$. Clearly, the parameter $\kappa \neq 0$ measures a distance away from the thermal equilibrium. If there is $\kappa = 0$, then the relation, Eq.(9), recovers the fluctuation-dissipation theorem. Thus, Eq.(9) introduces a generalized fluctuation-dissipation theorem, an energy-dependent relation of diffusion to friction, which exhibits a response of the interactions of the Brownian particle with its environment to its energy. Eq.(9) also accounts for an inhomogeneous or anomalous diffusion behavior in a system away from the equilibrium.

With the introduction of Eq.(9), from Eq.(8) it is found that the stationary-state solution becomes the power-law $\kappa$-distribution, exactly being Tsallis distribution [7] (here $\kappa$ is equal to $(1-q)$ in Tsallis distribution; do the same analogy hereinafter),

$$\rho_s(E) = Z^{-1}(1 - \kappa\beta E)^{1/\kappa}, \tag{10}$$

where the normalization constant is $Z = \iint dx dp (1 - \kappa\beta E(x,p))^{1/\kappa}$. In the limit $\kappa \to 0$, it recovers the Maxwell-Boltzmann distribution. In other words, the generalized fluctuation-dissipation theorem could account for the stochastic dynamics induced by the multiplicative noise to give rise to the stationary power-law $\kappa$-distribution. On the other hand, equivalently, Eq.(9) can be written as $D = m\gamma\beta^{-1}(Z\rho_s)^\kappa$, so we have

$$\frac{\partial}{\partial p}\left(D\frac{\partial \rho_s}{\partial p}\right) = Z^\kappa \cdot \frac{1}{\kappa+1} m\gamma\beta^{-1} \frac{\partial^2}{\partial p^2} \rho_s^{\kappa+1}. \tag{11}$$

Let $Z$ be associated with a constant $D_\kappa$ by $Z^\kappa = (\kappa+1)D_\kappa \beta(m\gamma)^{-1}$, from Eq. (11), the stationary F-P equation (5) reads

$$-\frac{p}{m}\frac{\partial \rho_s}{\partial x} + \frac{\partial}{\partial p}\left(\frac{dV}{dx} + \gamma p\right)\rho_s + D_\kappa \frac{\partial^2}{\partial p^2}\rho_s^{\kappa+1} = 0. \tag{12}$$

Accordingly, we receive a generalized Klein-Kramers equation for $\kappa \neq 0$,

$$\frac{\partial \rho}{\partial t} = -\frac{p}{m}\frac{\partial \rho}{\partial x} + \frac{\partial}{\partial p}\left(\frac{dV}{dx} + \gamma\, p\right)\rho + D_\kappa \frac{\partial^2 \rho^{\kappa+1}}{\partial p^2}, \tag{13}$$

whose stationary-state solution is exactly Tsallis distribution. This equation becomes the Klein-Kramers equation [21,22] if we take the limit $\kappa \to 0$.

As a particular example, $V(x)=0$, Biro and Jakovac presented a correlation strength of the noise in a momentum-space (i.e. the inhomogeneous diffusion coefficient, see Eq.(20) in [8]), $D(p) = D\left(1 + \frac{C}{D}p^2\right)$, by which they brought Tsallis velocity distribution from a linear Langevin equation. This diffusion coefficient is exactly a case of Eq.(9) when one takes $\kappa = -(2mC/D\beta)$, and with $D = m\gamma\beta^{-1}$ in terms of the fluctuation-dissipation theorem. The mechanism under the generalized fluctuation-dissipation theorem, Eq.(9), also well works for the plasma distribution in a superthermal radiation field, described by Hasegawa et al [2]. The theory of Hasegawa et al produced the famed power-law $\kappa$-distribution in plasma physics through the photon-induced Coulomb-field fluctuations. The velocity-space diffusion coefficient of the test particle is presented [2] by

$$D_\parallel(\upsilon) = D_\parallel^{eq(e)}\left(1 + \frac{D_\parallel^{NL(e)}}{D_\parallel^{eq(e)}}\right) = D_\parallel^{eq(e)}\left(1 + \frac{k_D^2|\gamma_0|^2}{3|\varepsilon(\omega_0,0)|^2 \ln \Lambda} \cdot \frac{\upsilon^2}{\upsilon_{Te}^2}\right), \tag{14}$$

which is found exactly to be a case of the generalized fluctuation-dissipation theorem, Eq.(9), without the potential $V(x)$, when one takes (the parameter $\kappa$ here is the reversal of that one in [2]),

$$\kappa = -\frac{2k_D^2|\gamma_D|^2}{3|\varepsilon(\omega_0,0)|^2 \ln \Lambda}, \tag{15}$$

where the mass of the test particle $m=1$, the mean square velocity $\upsilon_{Te}^2 = \beta^{-1}$, and the equilibrium diffusion coefficient $D_\parallel^{eq(e)} = \gamma\beta^{-1}$ in terms of the fluctuation-dissipation theorem have been taken into consideration.

We now consider another case of $f(E)$. Namely, if $f'(E) \approx \text{const.} > 0$ and the energy $E$ is high so that $\beta^{-1}$ can be neglected as compared with $f'(0)E$, we have

$f(E) \approx f'(0)E$, and then

$$D = m\gamma f'(0)E. \tag{16}$$

The stationary-state solution (8) becomes the power-law $\alpha$-distribution, with $\alpha = 1/f'(0)$, in such a form of

$$\rho_s(E) = A\left(\frac{E}{E_0}\right)^{-\alpha}, \tag{17}$$

where $A$ is the normalization constant; $E_0$ is an appointed reference energy to ensure the quantity in the bracket to be dimensionless. The power-law $\alpha$-distributions have been frequently observed and noted in some systems, such as e.g. the plasma distribution at high energy [2], the trapped ion collisions with heavy neutrals [10], the solar flares [14], the chemical and reaction-diffusion processes [15-17], or elsewhere.

The inhomogeneity in the correlation strength of the noise and the friction coefficient has not always to lead to Tsallis distribution. Clearly, the conditions under which Eq.(9) holds are particular cases of $f(E)$, and hence Eq.(9) is not the unique energy-dependent form of the relation of diffusion to friction. Theoretically, the stationary-state solution (8) exists if only $m\gamma/D \equiv f^{-1}(E)$ is an integrabel and positive function on the energy. This could account for the experimental observation recently on dust plasma [12] that not all systems with anomalous diffusion can be described by Tsallis distribution.

Finally, we consider the situation often paid attention to of the Brownian particle moving in a strong friction medium [20-22]. In this case, the coordinate undergoes a creeping motion, and the derivative on momentum may be eliminated approximately from the Langevin equation (2), leading to the following stochastic dynamical equations only for the coordinate,

$$\frac{dx}{dt} = -(m\gamma)^{-1}\frac{dV(x)}{dx} + (m\gamma)^{-1}\eta(x,t), \tag{18}$$

where the multiplicative noise satisfies

$$\langle\eta(x,t)\rangle = 0, \quad \langle\eta(x,t)\eta(x,t')\rangle = 2D(x)\delta(t-t'). \tag{19}$$

The time evolution of the corresponding reduced probability distribution function,

$\rho(x,t) = \int_{-\infty}^{\infty} dp \rho(x,p,t)$, is governed by the corresponding F-P equation,

$$\frac{\partial \rho}{\partial t} = (m\gamma)^{-1} \frac{\partial}{\partial x}\left(\frac{dV}{dx}\rho\right) + (m\gamma)^{-1} \frac{\partial}{\partial x}\left(D \frac{\partial \rho}{\partial x}\right). \tag{20}$$

It is ready to find that its stationary-state solution, $\rho_s(x)$, can be expressed as

$$\rho_s(x) \sim \exp\left(-\int \frac{1}{D} dV\right). \tag{21}$$

Now let us introduce the correlation strength of the noise, $D(x)$, as a function of the potential energy $V(x)$: $D = f(V)$. In the same way as Eq.(9), with a parameter $\kappa = -f'(0)$, we have

$$D = \beta^{-1}(1 - \kappa \beta V), \tag{22}$$

then it is derived that the stationary-state solution is the power-law $\kappa$-distribution, exactly given by Tsallis distribution:

$$\rho_s(x) = Z_\kappa^{-1}[1 - \kappa \beta V(x)]^{1/\kappa}, \tag{23}$$

where $Z_\kappa$ is a normalization constant, defined by $Z_\kappa = \int dx[1 - \kappa\beta V(x)]^{1/\kappa}$. Further, combining Eq.(22) with Eq.(23), we have $D = \beta^{-1}(Z_\kappa \rho_s)^\kappa$. And thus the stationary F-P equation is written as

$$(m\gamma)^{-1} \frac{\partial}{\partial x}\left(\frac{dV}{dx}\rho_s\right) + (\kappa+1)^{-1}(m\gamma\beta)^{-1} Z_\kappa^\kappa \frac{\partial^2 \rho_s^{\kappa+1}}{\partial x^2} = 0. \tag{24}$$

We let the normalization constant be associated with a constant, $D_\alpha$, by $Z_\kappa^\kappa = m\gamma\beta(\kappa+1)D_\alpha$, then we obtain a generalized Smoluchowski equation for $\kappa \neq 0$ as follows,

$$\frac{\partial \rho}{\partial t} = (m\gamma)^{-1} \frac{\partial}{\partial x}\left(\frac{dV}{dx}\rho\right) + D_\alpha \frac{\partial^2 \rho^{\kappa+1}}{\partial x^2}, \tag{25}$$

which has a power-law stationary solution, exactly being Tsallis distribution. Eq.(25) is the Smoluchowski equation [20-22] if one takes $\kappa=0$. Eq.(25) describes an anomalous diffusion behavior of the Brownian particle moving in a strong friction medium under the potential, with an inhomogeneous and energy-dependent diffusion coefficient, $(m\gamma\beta)^{-1}[1-\kappa\beta V(x)]$.

Eq.(22) determines one condition under which the generalized Smoluchowski equation has the power-law stationary solution, accounting for the stochastic dynamical origins of the power-law distributions as a result of the Langevin equation, Eq.(18) through the multiplicative noise, and showing a response of the diffusion to the potential. We notice that a thermostatistics of interacting particles in an over damped medium is investigated recently through a nonlinear F-P equation [9], which is a example of the above generalized Smoluchowski equation. As a particular case, i.e. the potential energy: $V(x)=\frac{1}{2}ax^2$, a general time-dependent solution of the generalized Smoluchowski equation is the Tsallis distribution [9],

In conclusion, the works in this letter provide a stochastic dynamical theory of the underlying power-law distributions induced by the multiplicative noise. We understand the conditions under which the types of power-law distribution are produced and how they are produced in systems away from equilibrium. The correlation strength of the noise and the friction coefficient in the two-variable Langevin equations are considered as inhomogeneous for coordinate and momentum. By seeking the stationary solutions of the corresponding Fokker-Planck equation, these power-law distributions induced by the noise are shown to be a result of that the relation of diffusion to friction depends on the energy, which reflects a response of the interactions of the particle with its environment to its energy. We hence derive the generalized fluctuation-dissipation theorems, which let us account for the microscopic dynamical origins of some stationary power-law distributions, including the $\kappa$-distribution in plasma physics, the $q$-distribution in Tsallis statistics, and those forms like $P(E) \sim E^{-\alpha}$. This leads to a generalized Klein-Kramers equation, and a generalized Smoluchowski equation for the particles moving in a strong friction medium, whose stationary-state solutions are exactly Tsallis distribution.